\documentstyle[12pt,aasms4, epsf]{article}

\def\Lsun{\hbox{\it L$_\odot$}}

\def\Msun{\hbox{\it M$_\odot$}}
\def\Minit{\hbox{\it M$_{\rm initial}$}}

\def\Myr{\hbox{\it Myr}}
\def\Gyr{\hbox{\it Gyr}}

\def\kms{\hbox{km$\,$s$^{-1}$}}

\def\AK{\hbox{\it A$_{\rm K}$}}
\def\K{\hbox{\it K}}

\def\simgr{\mathrel{\hbox{\rlap{\hbox{\lower4pt\hbox{$\sim$}}}\hbox{$>$}}}}

\makeatletter
\def\jnl@aj{AJ}
\ifx\revtex@jnl\jnl@aj\fi
\makeatother

\received{October 12, 1999}
\revised{REVISION DATE}
\accepted{January 10, 2000}
\journalid{VOL}{JOURNAL DATE}
\articleid{START PAGE}{END PAGE}
\paperid{MANUSCRIPT ID}
\cpright{TYPE}{YEAR}
\ccc{CODE}

\lefthead{Figer et al.}
\righthead{Sgr~A$^*$}
\begin{document}

\title{2 \micron\ Spectroscopy within 0\farcs3 of Sgr~A$^*$
\footnote{Data presented herein were obtained at the W.M. Keck Observatory, 
which is operated as a scientific partnership among the California Institute of Technology, 
the University of California and the National Aeronautics and Space 
Administration.  The Observatory was made possible by the
generous financial support of the W.M. Keck Foundation.}}

\author{
Donald F. Figer\altaffilmark{2}, 
E. E. Becklin\altaffilmark{3}, 
Ian S. McLean\altaffilmark{3}, \\
Andrea M. Gilbert\altaffilmark{4}, 
James R. Graham\altaffilmark{4},
James E. Larkin\altaffilmark{3}, \\
N. A. Levenson\altaffilmark{5}, 
Harry I. Teplitz\altaffilmark{6,}\altaffilmark{7},
Mavourneen K. Wilcox\altaffilmark{3}, \\
Mark Morris\altaffilmark{3}}

\authoremail{figer@stsci.edu}

\altaffiltext{2}{Space Telescope Science Institute, 
                  3700 San Martin Drive, Baltimore, MD 21218; figer@stsci.edu }
\altaffiltext{3}{Department of Physics and Astronomy, 
University of California, 
Los Angeles, 
Division of Astronomy, 
                 Los Angeles, CA, 90095-1562 }
\altaffiltext{4}{Department of Astronomy, University of California, Berkeley, 601 Campbell Hall,
                 Berkeley, CA, 94720-3411}
\altaffiltext{5}{Department of Physics and Astronomy, Johns Hopkins University,  
                 Baltimore, MD  21218}
\altaffiltext{6}{Laboratory for Astronomy and Solar Physics, Code 681, Goddard
Space Flight Center, Greenbelt MD 20771}
\altaffiltext{7}{NOAO Research Associate}

\begin{abstract}
We present moderate (R~$\approx$~2,700) and high resolution (R~$\approx$~22,000) 
2.0$-$2.4 \micron\ spectroscopy of the central 0.1 square arcseconds of the 
Galaxy obtained with NIRSPEC, the facility near-infrared spectrometer for the Keck II telescope.
The composite spectra do not have any features attributable to the brightest stars
in the central cluster, i.e.\ after background subtraction, W$_{\rm ^{12}CO(2-0)}$~$<$~2~\AA. 
This stringent limit leads us to conclude that 
the majority, if not all, of the stars are hotter than typical red giants. Coupled with 
previously reported photometry, we conclude that the sources are likely OB main sequence 
stars. 
In addition, the continuum slope in the composite spectrum is bluer than that of a red giant and is similar
to that of the nearby hot star, IRS16NW. It is unlikely that they 
are late-type giants stripped of their outer envelopes because such sources would be 
much fainter than those observed. Given their inferred youth ($\tau_{\rm age}$~$<$~20~\Myr), 
we suggest the possibility that the stars have formed within 0.1 pc of the supermassive black hole.
We find a newly-identified broad-line component (V$_{\rm FWHM}$ $\approx$ 1,000 \kms) to the 2.2178 \micron\
[\ion{Fe}{3}] line located within a few arcseconds of Sgr~A$^*$. A similar component is not seen in the
Br-$\gamma$ emission.
\end{abstract}


\section{Introduction}

Recent high-resolution near-infrared imaging reveals a tight cluster of at least a dozen
stellar sources projected within 0\farcs5 of the putative massive black hole in the Galactic Center\markcite{gen97,ghe98}
(Genzel et al.\ 1997; Ghez et al.\ 1998). Genzel et al.\ (1997) suggest that this cluster 
contains early-type stars with initial masses $\sim$15 to 20 \Msun.
While their low-resolution spectra (R $\approx$ 35) and photometry can be fit by early-type stars, they can also
be fit by much lower mass K giants of a few \Msun. The implications for
star formation near a massive black hole are heavily dependent on whether 
the stars are of early or late type. In
the former case, the stars are a few \Myr\ old and probably formed very near to the central
black hole. In the latter case, they are on the order of a \Gyr\ old and represent a
central concentration of the general old population seen throughout the Galactic
Center\markcite{ale99} (Alexander 1999). 
If stars have formed near the central black hole, then it is important to know
the physical properties of the gas there. The intensity of
recombination line emission can be used to constrain the gas density and ionizing environment near
the center, and the line width might be used to probe the central mass
to smaller size scales than the stellar velocity dispersions.

With these issues in mind, we obtained high resolution (R $\sim$ 22,400) and moderate
resolution (R $\sim$ 2,700) long-slit spectra of the central few arcseconds
of the Galaxy in the 2.0$-$2.4 \micron\ region ({\it K}-band). We present 
spectra of the combined light from the central stellar sources\markcite{gen97} (hereafter the ``S sources''; 
Genzel et al.\ 1997), nearby stars, and ionized gas. 

\section{Observations and Data Reduction}

The observations were obtained with NIRSPEC, the facility near-infrared spectrometer, on the
Keck II telescope\markcite{mcl98,mcl99} (McLean et al.\ 1998, 1999).
A log of observations is given in Table 1. The plate scale was measured by comparing the locations of the spectra of
IRS7, IRS16NW, IRS33E, and IRS33W to the positions in Eckart \& Genzel\markcite{eck97} (1997), giving
0\farcs14$\times$0\farcs20 per pixel in the spectral$\times$spatial directions of the high-resolution mode, 
and 0\farcs20$\times$0\farcs14 per pixel in the spectral$\times$spatial directions of the low-resolution mode
(note that the axes are flipped with respect to the camera in the two modes).
The slit viewing camera (SCAM)
was used to obtain images simultaneously with the spectra. We measured a plate scale for the SCAM of 
0\farcs18 per pixel by comparing the locations of IRS16NW, IRS16NE, IRS33E, and IRS33W to those given by
Eckart \& Genzel\markcite{eck97} (1997). From SCAM images, we estimate seeing (FWHM) of
0$\farcs$5 on 28 April 1999 and 0$\farcs$3 on 3 June 1999.
We chose to use the 3-pixel-wide slit (0\farcs43) in high-resolution mode and the 
2-pixel-wide slit (0\farcs39) in low-resolution mode in order to match the seeing. 

Quintuplet Star \#3, which is featureless in this
spectral region\markcite{fig98} (see Figure 1 in Figer et al.\ 1998), was observed 
as a telluric standard\markcite{mon94} (nomenclature from Moneti, Glass, \& Moorwood 1994). Arc
lamps containing Ar, Ne, Kr, and Xe, were observed to set the wavelength scale. A field
relatively devoid of stars 
(RA~17$^{\rm h}$~44$^{\rm m}$~49$\fs$8, DEC~$-$28$^{\arcdeg}$~54$^{\arcmin}$~6$\farcs$8~, J2000)
was observed to provide a dark current plus bias plus background image.
A quartz tungsten halogen lamp was observed to provide a ``flat'' image. 

All data reduction was accomplished using IRAF routines\footnote
{IRAF is distributed by the National Optical Astronomy Observatories,
which are operated by the Association of Universities for Research
in Astronomy, Inc., under cooperative agreement with the National
Science Foundation.}. Bad pixel removal, flat-fielding, and coadding of
object-sky frame pairs were performed to produce the final spectral
images. The spectra were extracted and then divided by
a similarly extracted spectrum of the telluric standard corrected for its apparent
spectral energy distribution\markcite{fig98} (Figer et al.\ 1998).
 
\section{Analysis and Results}

We have compared the flux in the spectrum to that expected from the stars
in the slit\markcite{wiz99} (Wizinowich et al.\ 1999), and 
the detected flux in the low-resolution spectrum confirms our pointing.  
The total background-subtracted flux from the combined low-resolution spectrum is \K\ = 12.8, in good agreement
with the sum of the previously reported fluxes for the sources, 
reported as \K\ = 12.6\markcite{gen97} (Genzel et al.\ 1997). 

Figure 1 shows the low resolution spectra for the S sources, IRS7, and IRS16NW.
The spectra demonstrate the very deep CO absorption in the spectrum of a 
cool star (IRS7) and the total lack of any similar
feature in that of a hot star (IRS16NW). 
The spectra for the S sources were extracted from synthetic apertures centered
1\farcs38 south of IRS16NW and 0\farcs75 wide to give
W$_{\rm ^{12}CO(2-0)}$~$\approx$~6$\pm$2~\AA. Figure 2 shows that the CO bandhead strength gradually increases
in apertures to the south of Sgr~A$^*$, and to the north of IRS16NW. For instance, 
W$_{\rm ^{12}CO(2-0)}$~$\approx$~11$\pm$2~\AA\ for
an aperture centered 0\farcs78 to the south of Sgr~A$^*$. 
A similar measurement for $\iota$ Cep (K0III), whose spectrum was taken from 
the Kleinmann \& Hall atlas\markcite{kle86}
(1986), gives W$_{\rm ^{12}CO(2-0)}$~=~11.5 \AA; note that later red giants would have deeper CO absorption. 
It appears, then, that the off-source spectrum is similar to that of a K giant. The total on-source 
continuum flux level (with background) is about 1.8 times the off-source flux level. So, if the collective spectrum for
the S sources is truly featureless, and the spectrum of the background mimics that of a K giant, then we
would expect to measure an equivalent width of 6~\AA\ in the combined spectrum, in good agreement
with our measurements. We find that no more than 30\% of the
light in the composite spectrum can come from K giants or later types. 
Note that this limit is conservatively 
determined by using the upper error limit
of W$_{\rm ^{12}CO(2-0)}$~=~8~\AA, for the combined light spectrum. 

The spectrum of the light from all sources in the central 0\farcs39 $\times$ 0\farcs85 
(EW $\times$ NS) suggests at least some
population of stars earlier than K0, and we find that the nearest blue supergiant, IRS16NW, does
not contribute significantly to the light falling in the aperture. 
The peak of the S sources is 1\farcs43
south of IRS16NW, and the point spread function (measured from IRS7) suggests that the light from 
IRS16NW should contribute $<$ 1\% of the total flux in the synthetic aperture. 

We are also interested in detecting possible emission-line flux from the stellar sources, in order to
constrain their nature, and from 
any gas associated with the black hole, in order to constrain the physical properties of material
near the black hole. The high resolution data are particularly useful for this purpose (see Table 1
and Figure 3).
We find three distinct components to the Br-$\gamma$ emission: 1) a faint
``zero-velocity'' component which is distributed throughout the low-resolution slit region, and is
not easily identified in Figure 3, 2) a very
bright, high-velocity, component associated with the ``mini-spiral,'' and 3) a near-zero-velocity component
near the position of Sgr~A$^*$ and extending 1\farcs5 to the north and 7\arcsec\ to the south. The first
component can also be seen in the H$_{2}$ (2.122 \micron) and \ion{He}{1} (2.058 \micron) lines, suggesting
that \ion{H}{1} and \ion{He}{1} gas are distributed in projection over the whole region, and are being ionized by the
ambient radiation field. The third component varies considerably along the north-south direction 
in peak location, 0 $\lesssim$ V$_{\rm LSR}$ $\lesssim$ +55 \kms,
and line width, 25~\kms~$\lesssim$~V$_{\rm FWHM}$~$\lesssim$~110~\kms, gradually increasing from south to north.
At the position of Sgr~A$^*$, the emission is centered at V$_{\rm LSR}$~=~+15~\kms, 
and V$_{\rm FWHM}$~=~80 \kms\ at the peak emission.
Within 2\arcsec\ of Sgr~A$^*$, the line center has a linear gradient of $+$18~\kms~arcsec$^{-1}$ (south-to-north),
and the line width has a peak value of +90~\kms\ at $-$0\farcs6 south of Sgr~A$^*$.
The only distinguishing characteristic of the emission at the location of Sgr~A$^*$ with respect
to surrounding regions is the local maximum in the line width which is 50\% greater than the
value $\approx$~1\arcsec\ in either direction. The otherwise nondescript appearance of the line
emission near Sgr~A$^*$ suggests that it has little to do with the black hole.
Forbidden iron-line emission can also be seen throughout the spectroscopic field.
There appears to be a broad [\ion{Fe}{3}] emission line near 2.2178 \micron\ 
with V$_{\rm FWHM}$ $\approx$ 1,000 \kms, peaking near the center of the spectroscopic field
shown in Figure 2; a similar broad feature is not seen in the Br-$\gamma$ line.
There is also a velocity component of
the [\ion{Fe}{3}] emission lines at 2.1451 \micron, 2.2178 \micron, 2.2420 \micron, and 2.3479 \micron\ which is
narrow and follows the pattern of the mini-cavity, the spatial distribution of which agrees well 
that described in Lutz, Krabbe, \& Genzel\markcite{lut93} (1993).

\section{Discussion}

The S sources span a range of brightness of 14.0~$\leq$~\K~$\leq$~16.0, implying a range of absolute
magnitudes of $-$3.2~$\leq$~M$_{\rm K}$~$\leq$~$-$1.1, assuming d=8000 pc\markcite{rei93} 
(Reid 1993) and \AK\ $\approx$ 2.7.
The absolute magnitudes match
those of O9V to B1V stars having 25~\Msun~$>$~\Minit~$>$~10~\Msun, and 
6~\Myr~$<$~$\tau_{\rm age}$~$<$~20~\Myr\markcite{mey94} (assuming the Geneva models, twice mass loss rate,
twice solar metallicity, and $\tau_{\rm age}$~$\sim$1~\Myr\ old; Meynet et al.\ 1994). 
The absolute magnitudes can also be fit by red giants of type K0III to
K3III with ages $\approx$~1~\Gyr. The early-type stars are most easily distinguished from the
red giants by the presence of CO absorption in the \K-band spectra of the latter. For instance,
the deepest CO bandhead in the \K-band (2.2935 \micron) for an early K giant 
has W$_{\rm ^{12}CO(2-0)}$~$\approx$~11~\AA.

The data suggest that the majority of the S sources in our synthetic aperture are hot stars
formed within 0.1 pc of the black hole. This is in agreement with Genzel et al.\markcite{gen97} (1997) and
Eckart, Ott, \& Genzel\markcite{eck99} (1999). It is unlikely that the observed stars are actually red giants minus their 
outer envelopes, such as might be produced via stellar 
collisions\markcite{lac82bai99} (Lacy, Townes, \& Hollenbach 1982; Bailey \& Davies 1999). 
Even a particularly luminous red giant ({\it L}~$\approx$~10$^4$~\Lsun) with
its envelope stripped to an extent consistent with the lack of strong Br-$\gamma$ absorption
would have \K~$>$~16, far too faint to be a likely candidate
for the S sources. 
It is also unlikely that they have formed outside of the center and have been transported inward via dynamical friction.
Using Figure 1 and equation 1 in Morris\markcite{mor93} (1993), we find that it would take 
longer than the lifetime of an O- or B-star to transport them into the center if they were formed 
more than 0.1 pc from the center.

The requirements for star formation so near the supermassive black hole are extreme.
Consider a protostellar clump of sufficient density to form an O-star near the black
hole. Such a clump would need to be very dense to be bound against tidal disruption:
\begin{equation}
\rho_{\rm clump}~\gtrsim~3.53~{{M} \over {R_{\rm GC}^3}}, 
\end{equation}
where {\it R$_{\rm GC}$} is the distance between the clump and the Galactic Center, 
and {\it M} is the enclosed mass within the orbit of the clump. Let's assume that 
the clump is as far away from the center as possible while
still allowing for dynamical friction to operate as described above, i.e., {\it R$_{\rm GC}$}~$\approx$~0.1~pc.
With M~=~2.6(10$^6$)~\Msun, we find, n$_{\rm clump}$~$\gtrsim$~4(10$^{11}$)~cm$^{-3}$. 
Since gas near the GC is presently at least 
5 orders of magnitude less dense than this, one must appeal to a much denser
environment during the formation epoch, or to an event which was exceptionally
strongly compressive, or to both.

As discussed by Morris, Ghez \& Becklin\markcite{mor99} (1999), the formation of stars this close to the
supermassive black hole would inevitably be accompanied by the
violent release of accretion energy with a total luminosity near
the Eddington limit of the black hole, since the black hole would
then be immersed in a relatively dense medium.  Indeed, this 
outpouring of energy may be required to compress the gas to 
densities sufficient to overcome the tidal forces.  
Given the challenge of forming stars in this tidally extreme 
environment, other possibilities might be considered.  

For example, have the masses of these stars been built up by 
stellar coalescence or by continuous accretion, making them much
older than we infer?  The calculations of Lee\markcite{lee96} (1996) suggest that 
this is unlikely, although Bonnell, Bate, \& Zinnecker\markcite{bon98} (1998) argue
that all stars with \Minit~$>$~10~\Msun\ form by coalescence in very young dense 
clusters; it remains to be seen if the steady-state conditions in the central parsec 
can mimic those in the formation epoch of a young cluster.
Or are these ``stars'' really more exotic 
objects such as compact stars with atmospheres acquired by passage
through a dense medium or by collisions with red giant stars\markcite{mor93} (Morris
1993)?  Perhaps they are stars powered by dark matter annihilation
in their interiors\markcite{sal89} (Salati \& Silk 1989), in which case they
would be much longer-lived than a star of comparable mass.  

In any of these cases, the explanation for the presence of these
stars would be exceedingly interesting, and continued investigation
well worthwhile. 

\section{Conclusions}
We find that about half a dozen of the stars projected within a few 
thousand AU of Sgr~A$^*$ have little, if any, CO absorption
in their {\it K}-band spectra, indicating that the stars are hot. Coupled with their brightnesses, we suggest
that the stars are OBV types, and therefore $<$~20~\Myr\ old. 
Given the lifetimes of such stars, it is improbable that they formed beyond 0.1 pc of the Galactic Center,
forcing us to consider the possibility that gas clumps having n~$\gtrsim$~10$^{11}$~cm$^{-3}$ can
exist within a few thousand AU of a supermassive black hole.

\acknowledgements

It is a pleasure to acknowledge the hard work of past and present members
of the NIRSPEC instrument team at UCLA: Maryanne Angliongto, Oddvar
Bendiksen, George Brims, Leah Buchholz, John Canfield, Kim Chin, Jonah
Hare, Fred Lacayanga, Samuel B. Larson, Tim Liu, Nick Magnone, Gunnar
Skulason, Michael Spencer, Jason Weiss and Woon Wong. In addition, we
thank the Keck Director Fred Chaffee, CARA instrument specialist Thomas A.
Bida, and all the CARA staff involved in the commissioning and integration
of NIRSPEC. We especially thank our Observing Assistants Joel Aycock, Gary
Puniwai, Charles Sorenson, Ron Quick and Wayne Wack for their support.
Finally, we thank Diane Gilmore of STScI for assisting in preparing the figures.

\clearpage
\begin{deluxetable}{rcrrrcr}
\small
\tablewidth{0pt}
\tablecaption{Log of Observations}
\tablehead{ 
\colhead{Resolution\tablenotemark{a}} & 
\colhead{Filter} & 
\colhead{Integ.} & 
\colhead{Frames} &  
\colhead{Slit Size} &
\colhead{PA} &
\colhead{Date} 
}
\startdata
22,400 & 1.561~\micron~$-$~2.312~\micron & 300 s. &  3 & 0$\farcs43\times 24\arcsec$ & 0$\arcdeg$ \& 90$\arcdeg$ & 28 April 1999 \nl
 2,700 &  1.996~\micron~$-$~2.382~\micron &  50 s. & 15 & 0\farcs$39\times 42$\arcsec  & 0$\arcdeg$ & 3 June 1999 \nl
\enddata
\tablenotetext{a}{The resolution is $\lambda$/$\Delta\lambda_{\rm FWHM}$, where $\Delta\lambda_{\rm FWHM}$
is the half-power line width of unresolved arc lamp lines.
The slit width was 2 pixels in low-resolution mode and 3 pixels in high-resolution mode.}
\tablecomments{All images were obtained using the multiple correlated read mode with
16 reads at the beginning and end of each integration.}
\end{deluxetable}

\newpage

\figurenum{1}
\figcaption[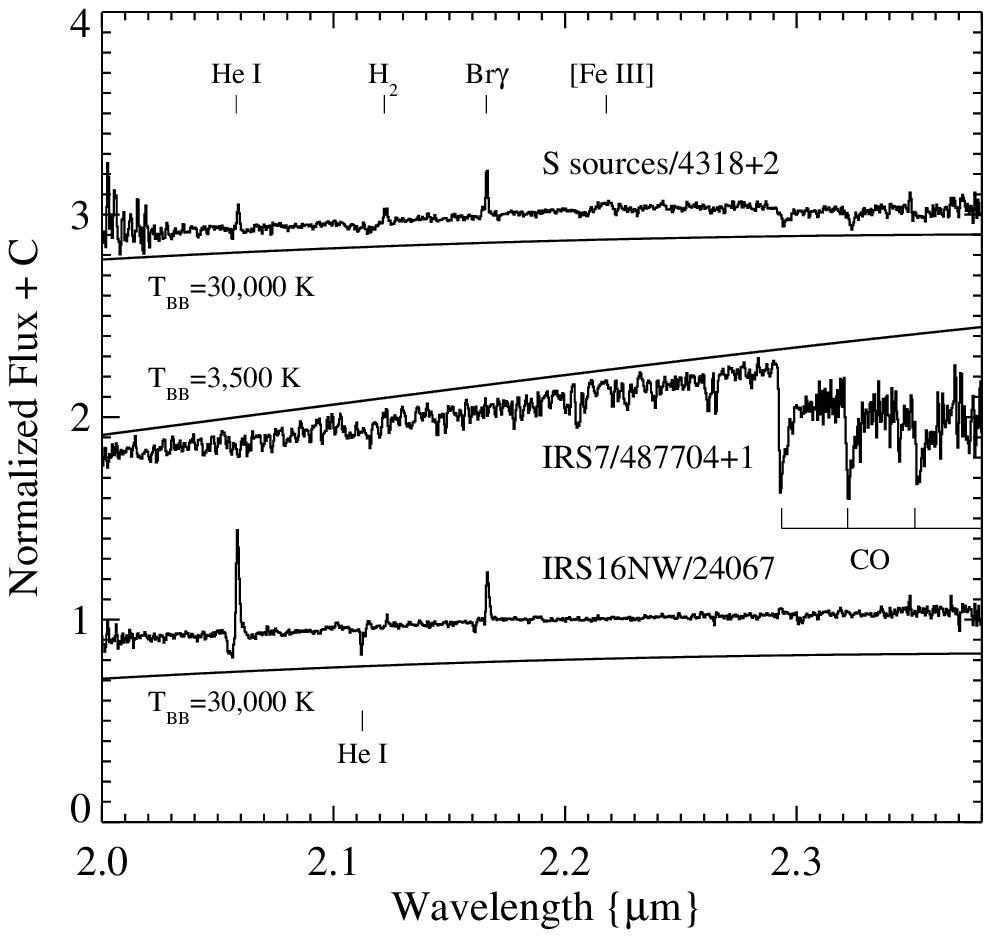]
{Low-resolution {\it K}-band spectra of the combined light from the S sources (S/N$\sim$120), 
IRS7 ($>$200), and IRS16NW (200). Reddened blackbody fits have been overplotted. 
We assume \AK~=~3.5 for IRS7 and 2.7 for the S sources and IRS16NW.
Normalization factors and constant offsets are given in the plot labels.}

\figurenum{2}
\figcaption[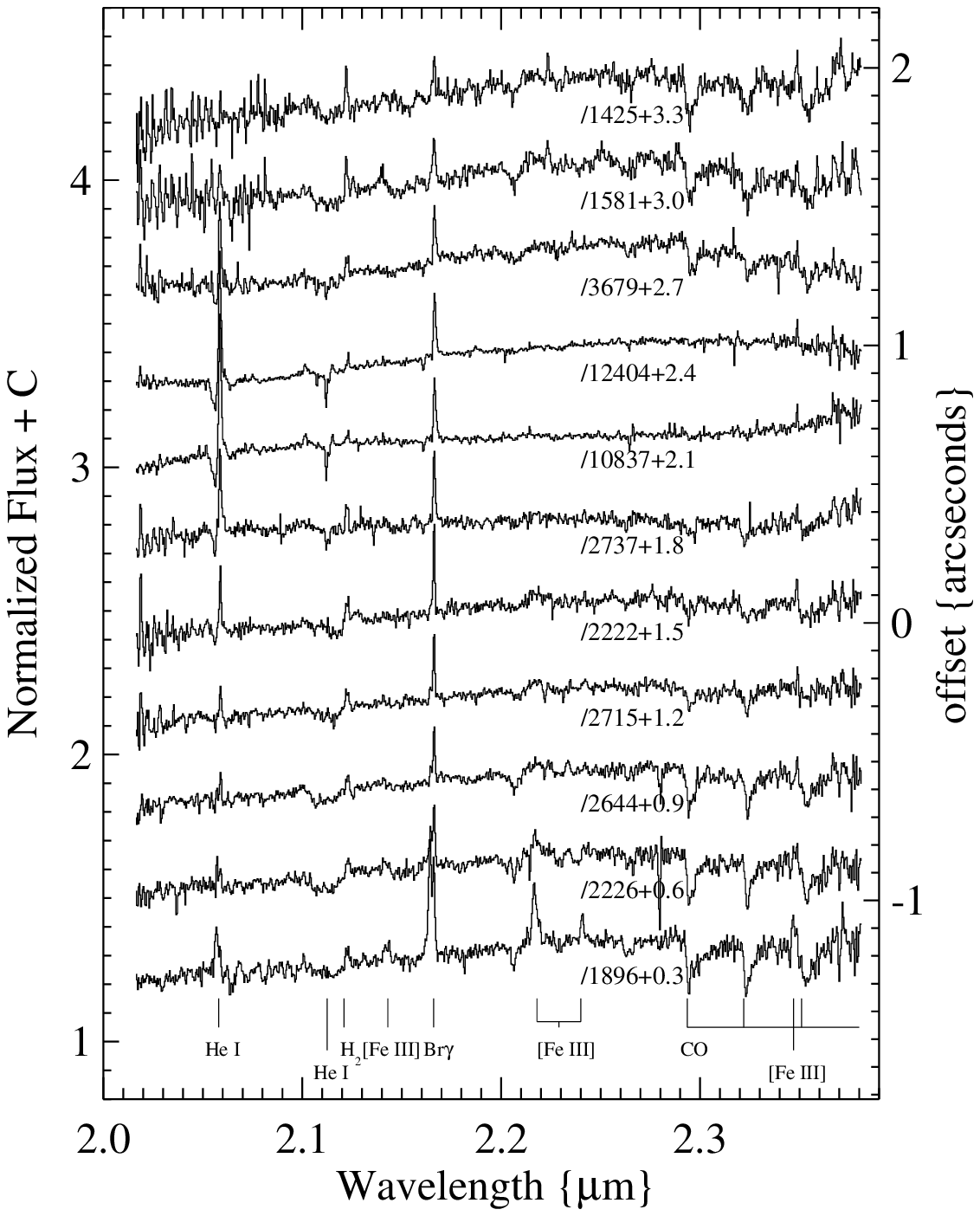]
{Spectra extracted from the low resolution data for synthetic apertures separated 
by 0\farcs36  in the north-south direction and offset 
with respect to Sgr~A$^*$. The y-axis gives location of the aperture with respect to Sgr~A$^*$ --- north is up.
Normalization factors and constant offsets are given in the plot labels.}

\figurenum{3}
\figcaption[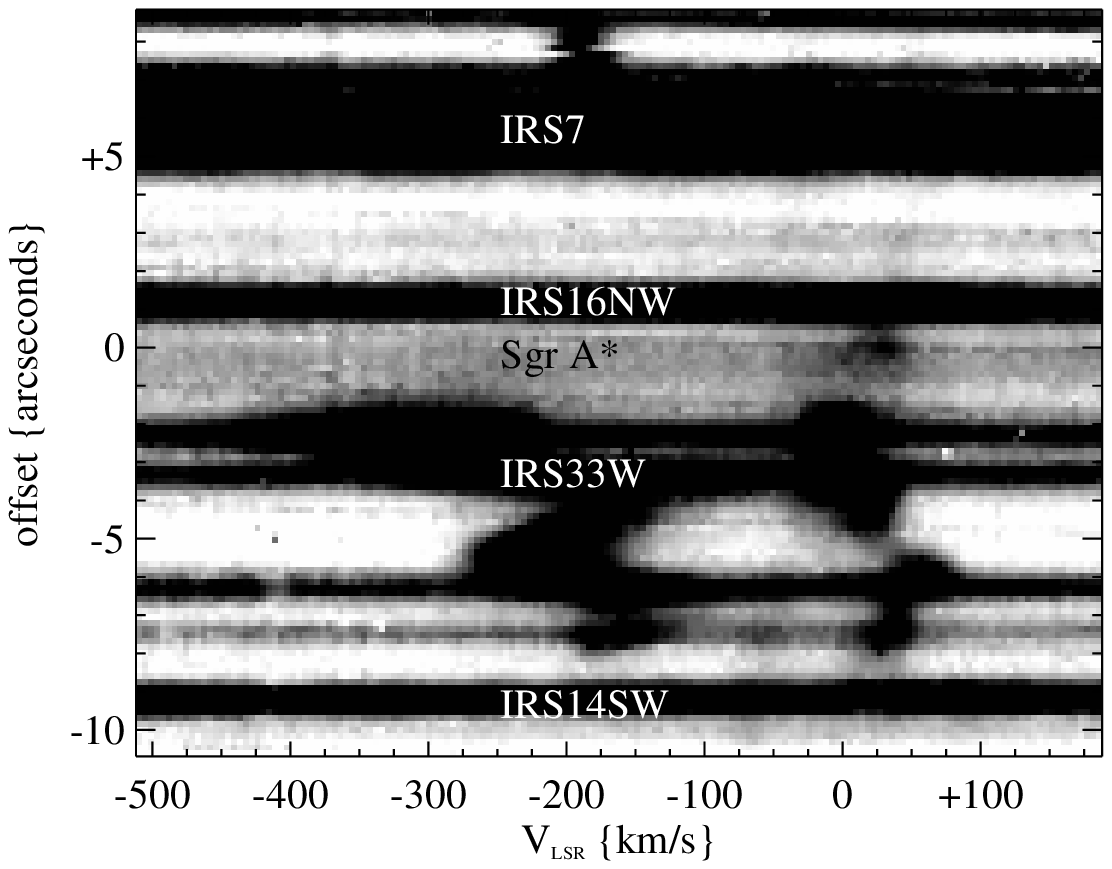]
{High resolution spectral image, displayed in inverted grayscale,
near the Br-$\gamma$ line. The slit orientation was north-south.
Objects discussed in the text are located at offsets of 0$\arcsec$ (Sgr A*), +1$\arcsec$ (IRS16NW), and
+6$\arcsec$ (IRS7).}

\newpage

\begin{figure}
\epsscale{1}
\hspace{3.75in}
\plotone{gcspectra.ps}
\hspace*{4.5in} 
\vskip .2in
Figure 1
\end{figure}

\begin{figure}
\hspace{0.8in}
\epsscale{.80}
\plotone{ndts03xxbf2d.ps}
\hspace*{4.5in} 
\vskip .2in
Figure 2
\end{figure}

\begin{figure}
\epsscale{.80}
\plotone{highres.ps}
\hspace*{2.2in} 
\vskip .2in
Figure 3
\end{figure}

\end{document}